# The Interplay of Learning, Analytics, and Artificial Intelligence in Education: A Vision for Hybrid Intelligence

Mutlu Cukurova
University College London, United Kingdom

## Abstract

This paper presents a multi-dimensional view of AI's role in learning and education, emphasizing the intricate interplay between AI, analytics, and the learning processes. Here, I challenge the prevalent narrow conceptualisation of AI as tools, as exemplified in generative AI tools, and argue for the importance of alternative conceptualisations of AI for achieving human-AI hybrid intelligence. I highlight the differences between human intelligence and artificial information processing, the importance of hybrid human-AI systems to extend human cognition, and posit that AI can also serve as an instrument for understanding human learning. Early learning sciences and AI in Education research (AIED), which saw AI as an analogy for human intelligence, have diverged from this perspective, prompting a need to rekindle this connection. The paper presents three unique conceptualisations of AI: the externalization of human cognition, the internalization of AI models to influence human mental models, and the extension of human cognition via tightly coupled human-AI hybrid intelligence systems. Examples from current research and practice are examined as instances of the three conceptualisations in education, highlighting the potential value and limitations of each conceptualisation for education, as well as the perils of overemphasis on externalising human cognition. The paper concludes with advocacy for a broader approach to AIED that goes beyond considerations on the design and development of AI, but also includes educating people about AI and innovating educational systems to remain relevant in an AI-ubiquitous world.

1. **Human Intelligence and Artificial Information Processing**

Artificial Intelligence (AI) is often defined as the simulation of intelligence in machines. Intelligence is a complex and multifaceted concept that encompasses several abilities. It does indeed include the capacity to learn, understand, reason, make decisions, and adapt to new situations (Norvig & Russell, 2010). It expands beyond commonly considered cognitive abilities, to include emotional and social components (Martinez-Miranda, & Aldea, 2005), acknowledging that intelligence is not just about how well one reasons, but also about how well one interacts with the world and others in it.

Intelligence is not only about what is certain, decontextualized, disembodied, tokenised and reduced to its parts, so that it is predictable and controlled. It is also about understanding things that are fluid. It is about the ability to live with, and survive despite uncertainty, and that what we are seeing as parts may indeed be wholes at another level. Today, most AI we see in educational research and practice considers AI as tools that have been developed to replace decision-making processes through analysis of big data, and prediction of the best value for a designated outcome variable, which is conveyed through a user interface. Intelligence, in my opinion, is more than that.

What do I mean by AI as tools? Let's have a look at it through the most prominent AI tool of today, OpenAI's ChatGPT[1]. Most state-of-the-art language models today including ChatGPT are based on a transformer architecture. During pre-training, a large-scale dataset of sentences is used as input to the transformer architecture. The inputs, for example masked-out words or paired sentences, are processed automatically and the neural network model is optimized to reconstruct the original text. First, an input is fed into the neural network, and it passes through the network's layers to produce an output. This process is known as the forward pass, where each layer's output is the input for the next layer, culminating in a final output from the network and it provides some predictions on masked words, next sentences, and so on. Since the actual masked words and next sentences in the original text are known, based on the differences between actual and model-predicted labels, a loss function is calculated. This function measures how far the network's prediction is from the actual result and backpropagation is used to minimize the loss by adjusting the weights of the network. This process of forward pass, loss calculation, backpropagation, and weight update is repeated over many iterations (or epochs) across the entire training dataset leading to the final pre-trained large language model. After the pre-training stage, Large Language Models (LLMs) are commonly fine-tuned to improve their performance. This is the subsequent process of refining the model on a smaller, more specific dataset to adapt it to a particular domain or task (e.g. through reinforcement learning with human feedback). They are also further prompt-tuned which involves optimizing the input prompts to guide the pre-trained model's behaviour on specific tasks without actually changing the model's parameters. After these large language models are trained, they are used to help or replace decision-making processes for the particular task of text generation and their predictions are conveyed through a user interface.

To a certain extent, this is a simplified description, but it broadly covers the essence of LLMs' training process without getting into the details (e.g. attention mechanisms (Vaswani *et al*., 2017)). This approach leads to the current best performing generative AI models and tools of the day. However, due to their non-transparent nature (i.e. they are black-box systems trained on vast amounts of data and their internal workings are complex), lack of understanding of the real-world contexts, and tendency to generate incorrect information, their value for certain learning tasks might be limited. They also lack clear, reliable, and valid measures of success in educational contexts which makes it difficult for performance evaluations in real-world implementations (Chang *et al*., 2024). This is not to say they do not have any value for educational purposes since they can still be very valuable for certain productivity gain tasks (e.g. generating initial draft content to be reviewed) and diagnosis purposes (e.g. predicting particular language issues with relatively high accuracy to support teachers in their prioritisation of interventions for their learners).

---

[1] https://chatgpt.com/chat

Many researchers and practitioners of AIED already know these in detail, so the readers might be wondering why we need to keep reminding ourselves about the gist of how these models work, and why this is important in this paper on hybrid intelligence. It matters because when we think about AI in these terms it does not cover all the terms we would use when we are describing human intelligence, it sounds more like artificial information processing, rather than artificial intelligence. Information processing is an important aspect of intelligence, but diminishing the whole concept to it would be a mistake. This is one of the reasons, LLMs are sometimes referred to as 'stochastic parrots' (Bender *et al*., 2021). Stochastic in that they generate content based on probability analysis, and parrot because they do not appear to have an understanding of the meaning of anything they generate. LLMs are good language models, but they are not models of the real-world in which the meaning is situated. Yet, this is not a competition between human intelligence and artificial information processing. We do not necessarily need more replications of human intelligence in machines for hybrid intelligence. Humans are very good at many things that today's AI is still pretty poor at, and AI is good at some others. Machines are much ahead of humans on some variables like computing floating point arithmetic, yet way behind on others like cognitive flexibility and long-term planning in unusual situations. This is not to say that humans are more intelligent than machines, or vice versa, we are differently abled.

Recognition of these differences in abilities provides a strong argument for the value of hybrid intelligence systems which are tightly coupled human-AI systems where both entities interact smoothly and dynamically, leveraging their respective strengths. In educational contexts, hybrid intelligence systems can significantly enhance human competence development by combining human cognitive flexibility, reflective long-term planning, and real-world contextual understanding with AI's data processing capabilities and analytics. If the relationship holds (which also requires humans to develop certain competencies as briefly covered in section 9.1), hybrid intelligence systems would be invincible. This relationship, as will be detailed in later sections, not only holds the potential to achieve productivity gains in human tasks but also help in avoiding human cognitive atrophy, and its potential amplification, by ensuring that humans remain engaged in critical thinking and decision-making processes.

## 2. AIED and The Direction of Research towards AI as an Applied Tool

When we start thinking about intelligence in these broad terms, it becomes clear that considering AI as an applied tool is only one part of a much bigger picture. AI can also be considered as a method for understanding human intelligence and learning, an opportunity to understand the differences and similarities between humans and artificial information processing. In this sense, studying human learning and studying AI are naturally intertwined. Interestingly, this way of thinking was indeed prevalent in the early roots of the learning sciences and AIED communities. Yet, it seems that we are losing this connection. In a review piece mapping publications from past and recent AIED conference proceedings and International Journal of AI in Education manuscripts on the coordinates of using AI as an applied tool vs AI as an analogy to human intelligence to study learning processes; Rismanchina and Doroudi (2023) showed that although early publications had well-distributed contributions on these coordinates, there was only one single publication in 2021 focusing on the use of AI as an analogy to human intelligence for studying learning. Why is that?

AI in Education can also be conceptualised to externalize, to be internalised, or extend human cognition (Cukurova, 2019). As the first conceptualisation, in the externalization of cognition, certain human tasks are defined, modelled and replaced by AI as a tool. In the second conceptualisation, AI models can be used to help humans change their representations of thought, through the internalization of these models. At last, AI models can be used to extend human cognition as part of tightly coupled human and AI hybrid intelligence systems. It is important to note that in such systems, changes in both agents are expected to be observed through their interactions and the whole emergent intelligence is synergistic, that is, it is expected to be more than the sum of each agent's intelligence, both human and artificial (Cukurova, 2019).

If we try to map these conceptualisations on the coordinates of Schneiderman's human control and agency versus automation through AI (Shneiderman, 2020), perhaps most traditional educational technology could be considered to have a low allowance for human agency and low automation built into them (see the AIED-HCD conceptual framework in Figure 1 below). With the initial proliferation of AIED research field, many researchers had the ambitious goal of creating systems that are as perceptive as human educators through the automation of certain tutoring behaviours, which led to significant developments in intelligent tutoring systems (ITSs). A similar trend in increased popularity of the externalisation conception can also be observed in most of today's LLMs and generative AI tools.

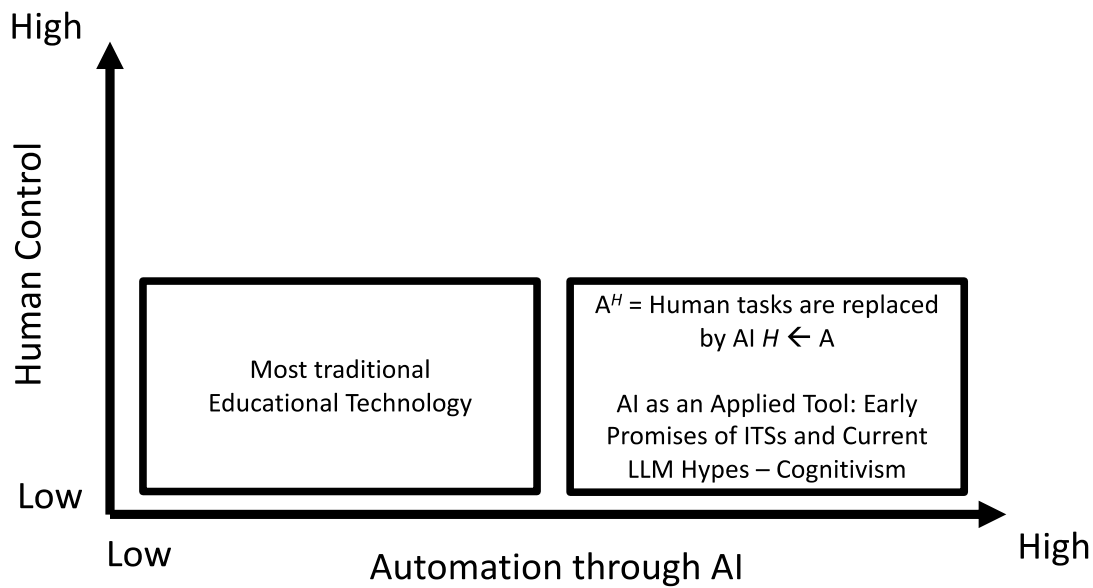

**Figure 1.** The AIED-HCD conceptual framework for human-AI interaction in education for human competence development – Human tasks are replaced by AI.

Mainly driven by the cognitivist approaches that consider learning only as an intracranial activity of information processing, ITSs have the main goal of adapting to the individual levels of mastery and needs of each student, tailoring the content, pacing, and feedback accordingly. The field of AIED has multiple examples of early successful ITSs including Carnegie Learning (Ritter *et al.*, 2015), Duolingo (Von Ahn,

2013), and Assistments (Heffernan, & Heffernan, 2014) and similar tutors are also emerging with the use of LLMs in educational chat interfaces (e.g. Khanmigo). It is important to highlight that perhaps one of the reasons for the success of these systems compared to relatively less successful ones in the field is that they have engaged with end users (e.g. teachers, students, and school admins) and taken the dynamics of education systems, school environments, and classrooms very seriously early on. For Carnegie Learning, for instance, there are physical course books and other resources in addition to the tutoring system itself. There is significant guidance, advice, and support for how teachers and students should interact ahead of any use of the tutoring systems, as well as while they are being used. So, there is significant human agency support, both for the teachers and the students, at the deployment phase of these ITS examples even the tools themselves have pedagogical action externalisation and automation at the system levels.

ITSs have traditionally focused on pedagogical task automation in digital environments. However, these systems can also work with data from physical spaces and can focus on a range of affective, metacognitive, and engagement tasks. For instance, an example from Kawamura *et al.* (2022) is using multimodal data to detect students' engagement states while working with an ITS. The system processes data from students' heart rate, seat pressure, and facial recognition to model students' level of awakeness and aims to provide suggestions for rest or adjusts the content or feedback accordingly.

These systems are also not limited to traditional human-generated educational content but can be delivered with AI-generated synthetic media. Recent experimental research investigating the potential of using AI-generated synthetic video to create viable educational content for ITSs, no significant differences were observed in learning gains and learner experience between the two conditions of students learning from a recorded human lecture vs. from an AI-generated synthetic media that delivers the same content (Leiker *et al.*, 2023). Admittedly, this is from a relatively small sample of 83 adult learners in an explorative study, but these models are getting better by the day, and the promising results justify further large-scale explorations of their potential.

Such tutoring systems, that externalise particular pedagogical tasks to model and automate their support, have been a significant area of research in AIED and they work very well for various domains and knowledge acquisition tasks. More long-term evaluation studies and RCTs with larger sample sizes and better-controlled conditions in the real-world would always be welcome but there is indeed good evidence both at the individual studies level including but not limited to SQL-Tutor (Mitrovic, & Ohlsson 1999), ALEKS (Craig *et al.* 2013), Cognitive Tutor (Pane *et al.* 2014), ASSISTments (Koedinger *et al.* 2010) and also at the meta-reviews level. For instance, VanLehn (2011) found that the effectiveness of intelligent tutoring systems were nearly as effective as average human tutors; Ma *et al.* (2014) found similar results both when compared to no tutoring or to large group human-tutor instruction; Pane *et al.* (2014) found evidence of the relative effectiveness of online tutors over conventional teaching; in Kulik & Fletcher (2016)'s work, the median effect was observed raising test scores 0.66 standard deviations over conventional levels, or from the 50th to the 75th percentile and du Boulay, B. (2016) summarised some metareviews in his work and showed that these systems achieve positive results in the delivery of knowledge acquisition particularly for the subjects of Maths, Language learning, and Algebra.

## 3. Research evidence on effectiveness versus real-world impact of AI in Education

Considering these systems are not necessarily new, and the evidence about their effectiveness is not necessarily new, one question to ponder upon as a community is why they are still not prevalent in mainstream education?

There are various reasons why this is the case and each one of these reasons would probably require a paper on its own to be discussed in detail. But, let me attempt to mention a few reasons, from my point of view, briefly here. First, there are numerous factors influencing the adoption and use of AI in education that are broader than the effectiveness of the specific AI technology. These include but are not limited to policy landscape, institutional governance, pedagogical culture, technological infrastructure, and social support mechanisms provided to teachers. For instance, in our recent work looking at the factors influencing teachers' adoption of AI in schools with about 800 schoolteachers (Cukurova *et al.,* 2023), we observed that although AI-tool related factors were indeed important, they were not necessarily the most important factors influencing the teachers' engagement with AI in schools. Not generating any additional workload, teachers' knowledge of, and confidence in using AI, increasing teacher ownership, generating support mechanisms for help when needed, and assuring that ethical issues are minimised, were also essential for the adoption of AI in schools. So let us never forget that the tools we are working on are not only closed engineering systems but are part of a large socio-technical ecosystem, and many factors will influence their adoption and effectiveness.

Second, education with fully automated systems that externalise human cognition to deliver educational practice can be argued to dehumanise learning. When AI in education is considered in a narrow sense, as lonely individual learners working on their own with an AI system, this might indeed lead learners to prioritise information gathering and declarative knowledge acquisition over tacit knowledge and wisdom which comes through rich experiences in the real world. Particularly, if these systems are considered as a replacement for human interactions with each other, and with the real world, then the knowledge that comes through experience and practical acquisition of an embodied skill can be replaced with tokens of representations that are far off the actual construct. Learning is not only about absorbing information and education is not only about learning. These are also about developing social competence, emotional intelligence, and various metacognitive abilities in real-world interactions as well as serving to other societal needs. Fully automated systems are unlikely to deliver such experiences as a whole in the near future, even if it is ever possible. However, these technologies can also be used as opportunities for increasing students' interactions with adults, and with each other, in affluent, private schools (See for instance, Alpha School, Austin). So, the impact of AI in education is not only dependent upon the AI tool itself, but the particular learning design and the instructional design in which they are embedded.

Third, a significant amount of work is still needed to address socio-psychological barriers to the use of AI in educational contexts. Students need to be motivated enough to engage with AI tools in the first place, yet only about 5% of them manage to engage with educational resources long enough on their own to get statistically significant benefits (Weatherholtz *et al*., 2022). In addition, teachers and learners tend to have confirmation biases, and unrealistic expectations from AI (Nazaretysky *et al*., 2021). Previous research showed that when people are presented with content framed as coming from AI, they tend to judge it as less credible and trustworthy compared to the same content framed as products of educational psychology or neuroscience (Cukurova *et al*., 2020). Similar results are now emerging, when AI-generated content or feedback is presented to teachers and students they tend to judge its quality lower and trust it less if they know that it is AI-generated (Nazaretsky *et al*., 2024). There is emerging work focusing on how we measure and gain the trust of teachers and learners in AI, but considerably more research is needed in this space to optimise their trust (e.g. Nazaretsky *et al*., 2022).

Fourth, particularly with the goal of keeping the talk focused on AI design and development rather than opening it up to broader ecosystem level issues, one of the most significant limitations of these AI tools that externalize tasks to automate them is that they are commonly built on computational cognitive models that overlook the intricacies of socio-cultural learning occurring beyond an individual's mind (Doroudi, 2023). For many learning scientists who are aligned with constructivist learning theories, this approach is considered too simple to represent the complexities of the kind of learning they are interested in.

### 4. Alternative Conceptualisations of Artificial Intelligence

In this regard, another conceptualisation of AI can provide significant opportunities. AI can also be conceptualised as computational models of complex learning phenomena for humans to internalize and change their representation of thought (Kent *et al*., 2021). These lead to relatively low automation systems that allow high human agency and control (see the AIED-HCD conceptual framework in Figure 2 below).

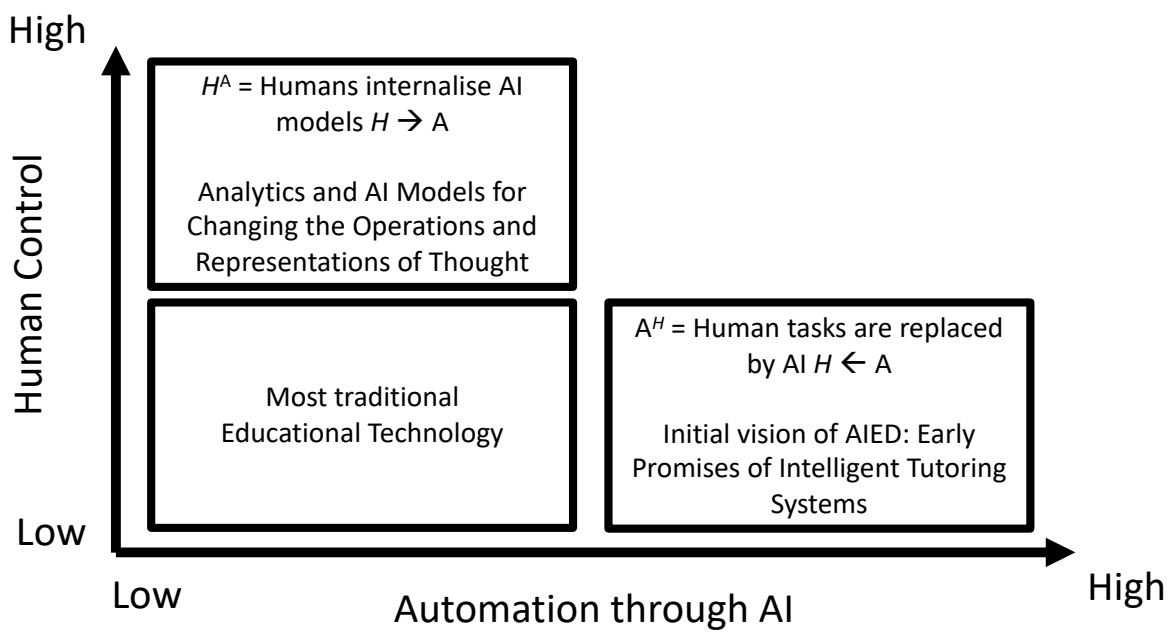

**Figure 2.** The AIED-HCD conceptual framework for human-AI interaction in education for human competence development – Humans internalise AI models.

In learning sciences literature, it is well established that learning may be facilitated with mental models (Johnson-Laird, 1983). Mental models help us explain and predict how people interact with the world, and how they explain, understand, solve anticipated events, and communicate. While mental models are internal structures, they can be exteriorised when triggered by interaction (Bransford, Brown, & Cocking, 2000). The assumption in this conceptualisation of AI is that computational models can offer an externalisation trigger for mental models. By that, they can serve as a learning affordance, and the learning outcomes can be observed through changes in learners' language, and their ability to explain, predict or diagnose phenomena (Kent *et al.*, 2021). So, although everyone comes to a learning situation with their own mental model of what success looks like, as well as the models that learners hold about themselves (Byrne, 1992), we can use data to model what success looks like with computational approaches and present these models back to people and trigger them to refine their mental models.

This conceptualisation allows more opportunities for researchers interested in understanding and designing learning environments through the lens of constructivist and socio-cultural learning perspectives. It allows opportunities to focus on AI models to help develop learner competence through rich learning experiences and reflecting on these experiences with the help of computational models.

In my research, we have been mainly focusing on such open-ended learning environments and trying to design analytics and AI models that would support teachers and learners in these constructivist contexts. For instance, we were investigating students who were engaged in solving open-ended design problems (e.g. Cukurova *et al.*, 2016) and have been collecting data through 2D and 3D video for face and gesture analysis, tracking the speed and distance between their hands, logging their physical computing kit interactions, their mobile tool reflections and notes, as well as their self-declared emotions (Spikol *et al.*, 2018).

If we take the first conceptualisation of AI, we can use these kinds of data to build machine learning classifications of success in these environments. First, using these different modalities of data, one can calculate various input measures as independent variables. For instance about the group itself such as the number of faces looking at the screen, their mean distances, their hand movement in space and speed, their audio features, gaze directions, emotional states etc., but also from the context such as the amount of time each group spends on different phases of the collaborative problem-solving process such as ideation (problem scoping), activities to solve the problem, and reflection on potential solutions. Then, you can ask researchers/teachers/experts to use existing learning sciences frameworks and evaluation metrics to label groups' competence while watching these interaction videos and meticulously labelling them (e.g. Cukurova *et al.*, 2018).

This leads to having multimodal measures of potentially relevant variables as input and labelled competencies of groups as the ground truth output. Then it becomes a machine learning problem of building reliable and accurate classification and regression models for the tasks of predicting competence classes or scores (e.g. Spikol *et al.*, 2018). In order to see the value of different modalities of data we also train models with various features removed, to often find out that the best results are achieved when multimodal data is used compared to unimodal predictions (Cukurova, Giannakos, & Martinez-Maldonado, 2020).

These models at this stage still have significant technical problems, and they are usually prototypical, rather than reliable tools for immediate real-world use. Regardless, often time the ultimate goal of such prediction models is to generate some kind of a dashboard for teachers and learners to directly intervene in the practice. For instance, Aslan *et al.*, (2019) used a Multimodal Learning Analytics (MMLA) tool with a dashboard to provide teachers with help to prioritise and structure their interactions with students. Their results indicate that when teachers are using the dashboard, they spend statistically significantly less amount of time in close monitoring actions and more amount of time in scaffolding activities. Students also appear to spend less time in bored states, and there can be a positive impact on their learning gains.

5. **The Challenge of Using AI as a Tool to Directly Intervene in Teaching and Learning**

However, the use of AI that directly intervenes in the practice of teaching and learning in such constructivist learning environments has significant challenges. These issues broadly relate to threatened human agency, the challenges of predictions in social contexts and the normativity issue of not being able to decide what is actually, or eventually, good or bad in a complex social learning situation (summarized in figure 3 below). Addressing these issues would require a stronger alignment between human values and AI goals which is a significant challenge highlighted by various other scholars (e.g. Russell, (2019); Christian, (2021)).

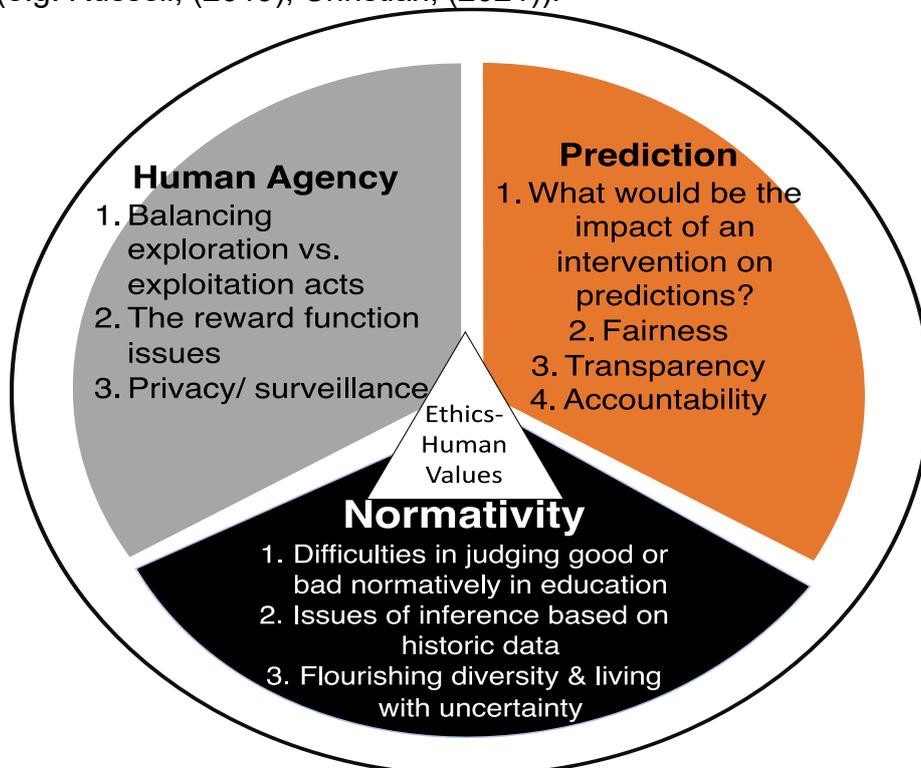

**Figure 3.** Main categories of issues related to using AI that directly intervenes in the practice of teaching and learning.

Of course, prediction issues also include well-documented challenges of algorithmic bias, transparency, and accountability of AI tools (e.g. Baker & Hawn, 2022), but I think the concern is even bigger than these. The issue is that in the design and development of these tools, we are used to doing engineering work where we can in effect see how the gears mesh to understand how things work. Whereas I am not sure if it is always possible to make a mathematical narrative or a model to explain, or predict, how a complex social system behaves. Is it always possible to explain or predict all aspects of human learning and human competence development? AI was coined as a term in the Dartmouth College summer school proposal in 1956 based on the conjecture that every aspect of learning or any other feature of intelligence can in principle be so precisely described that a machine can be made to simulate it (McCarthy, Minsky, Rochester, & Shannon, 1955). Since then, we somehow live under the impression that if only we could find them, there would be formulas and models to somehow predict all aspects of human learning. Yet, maybe to find out that to develop what such learning is, we just have to go through the same irreducible steps as the system itself. Maybe some aspects of learning just come through the slow experience of living those learning experiences. This itself makes the time spent on them more meaningful in the sense that, we just cannot jump ahead to get the answer with a prediction telling us what would be the most productive next step to take in these complex socio-cultural learning environments.

On the other hand, if we take the second conceptualisation of AI, as computational models for humans to internalize; these can be considered as opportunities to describe the learning processes in more precision, rather than aiming for "the potentially impossible task" of prescribing acts for the future based on predictions in a complex social learning process. In this sense, AI models become opportunities for thinking about learning. As Seymour Papert (2005) famously noted "You can't think about thinking without thinking about thinking about something." This suggests that to engage in the process of thinking, one must have something specific to think about. It underscores the importance of the objects of our thoughts in shaping the processes and pathways through which our thinking unfolds. In this sense, AI models can be objects to think about human learning. In learning analytics research, we often utilise this idea of "from clicks to constructs" (e.g. Knight, & Buckingham-Shum, 2017) aiming to describe precisely the relationship between the digital traces of data collected and the educationally meaningful constructs we are interested in supporting. Below is an example focuses on the concept of collaboration, moving from digital traces of audio data and video data processed with computer vision to model group interaction behaviours and their connections to outcome measures of shared understanding, satisfaction, and product quality (Zhou, Suraworachet, & Cukurova, 2024).

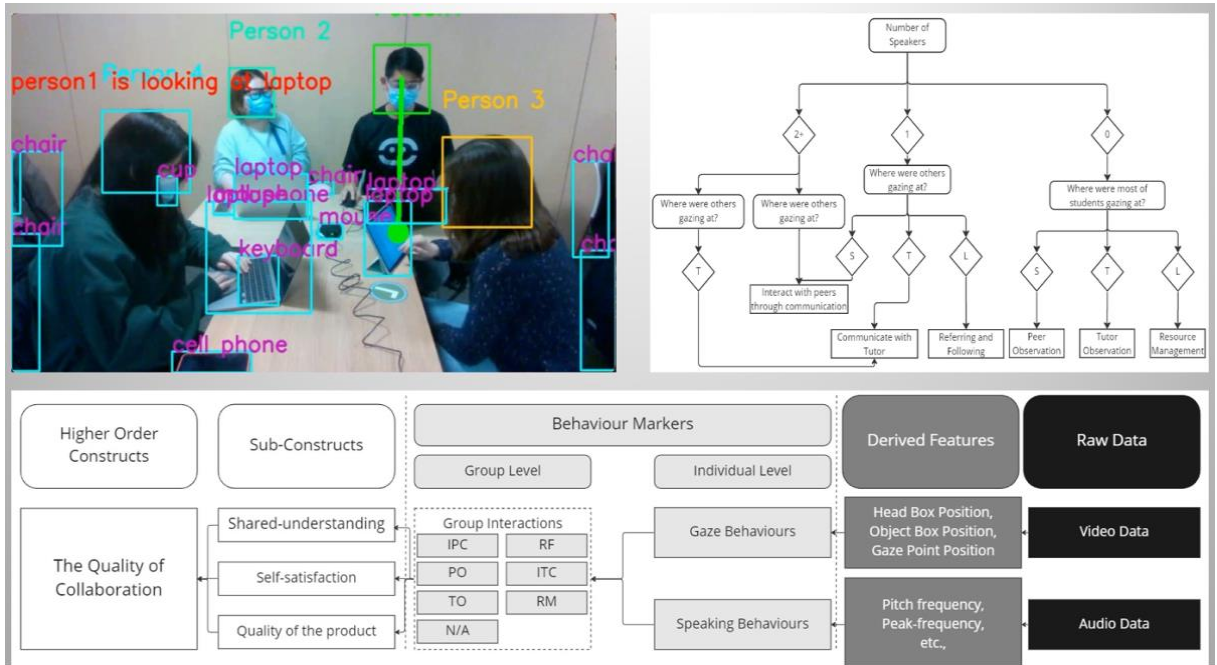

**Figure 4.** An example of the transitioning from digital traces to constructs of collaboration.

Sometimes, such models can give the impression that our attempts are for, or only, to build prediction models for prescriptive suggestions accordingly. If such an impression is taken forward, as the constructs are broken into small measurable components, more mechanistic measures of *how much, how fast,* or *how precisely* a component is completed may be prioritised. In the collaboration process model for example, as if one keeps eye contact for one more second, starts speaking with a peer one second earlier, clicks the resource button faster, or attempts to answer a question one more time, they would develop the expected competence in collaboration. For certain learning experiences, more and faster completion of a component, or even the more precise yet not flexible completion of it, may not lead to improvements in the quality of learning. Such measures are great for measuring the capabilities of machines, but not for all aspects of human learning. Taking a more modest goal of using AI models for describing these learning processes in more detail and precision can still allow significant opportunities for researchers, teachers, and learners.

## 6. Value of Making Lived Experiences Visible to End Users

For end users the models provide, specific and precise feedback opportunities to improve their awareness of the lived experiences and keep them motivated to engage with relevant activities in the future. Building on the same example, for instance, visualisations of speech time percentages and the types of group interactions each group can be used as feedback after group interactions. Similarly, different types of group interactions students engage with during the timeline of their activity and relevant feedback based on these interactions for self-awareness, reflection, and motivation for future interactions can be created and sent to students (e.g. Zhou *et al*., 2021) and teachers (Pozdniakov *et al*., 2022). In my view, these models are tools for making sense of the phenomena being modelled and sharing this sense with the end users, but they are not necessarily the mirrors of the reality representing the complex process of collaborative learning itself.

This recognition of the limitations of the AI models also provides opportunities for more realistic interactions between the model outputs and end users. For instance, we have been working on models that detect speech in groups, transcribe it from speech to text, and label the text with groups' challenge moments in their collaborative discourse (Suraworachet *et al.*, 2024). Then these detected challenge moments are also sent back to students as feedback with visualisations and further explanations based on the threshold values on certain aspects and diagnosed challenges, building upon the human interpretation of these values and providing suggestions on how to address these challenges in their next group activities. Rather than assuming what is being modelled and detected as the reality, we also provide students with their own transcriptions, as well as the details on how these models are built, and their episode level and sentence level predictions. So, they see the sense-making process we use in detail, and interpret it themselves when such feedback is valuable, or when it is safe to be ignored.

When evaluating the value of AI models with this conceptualisation for feedback to end users, it is always interesting to observe what meaning learners and teachers generate from the visible information from these models, what actions they take based on such an understanding, what is the accountability of this understanding, and what are their concerns related to their values and moral considerations (Erickson, & Kellogg, 2000). End users are often aware of the limitations of these models and agree about the incomplete nature of them. However, they still find them valuable to increase their awareness of their own learning activities, and also awareness of others' behaviours in these socio-constructivist learning environments (Zhou et al., 2021). The accountability of this awareness tends to influence both their motivation and engagement with the learning experience, as well as their regulation (including co-regulation and socially shared regulation (Hadwin, Järvelä, & Miller, 2011). Therefore, making the lived experiences of learning more visible and explicit with computational models still has significant value for teachers and learners regardless of their potential for accurate future predictions with prescriptive suggestions.

### 7. Value of Contributing to Learning Sciences Literature

Another value of this conceptualisation of AI as models for making sense of the phenomena being modelled is that they enable opportunities for clarification and communication of researchers' concepts in a more detailed, precise and formal language; generating potential insights into complex learning processes to advance learning theory (Giannakos, & Cukurova, 2023). For instance, in our recent work, we have been using five-channel multimodal data to make sense of Collaborative problem-solving processes. Instead of looking at each group's collaborative problem-solving activities separately, we merged all sequences from all groups in our dataset for a given task, then looked at clusters of patterns emerging using optimal matching algorithms and Ward's clustering, identifying three different clusters of CPS patterns in multichannel data streams at a granularity level that is not possible with traditional statistics. We then looked at the transitional and structural differences of each CPS pattern type with Hidden Markov Models and Epistemic Network Analysis to discover that they are associated with different performance outcomes (Ouyang, Xu, & Cukurova, 2023).

In turn, such insights can provide opportunities for learning theory to be further improved (Doroudi, 2023). A recently published literature review investigating the relationship between learning theories and models in MMLA research indicated that such models have the potential to contribute to learning theory (Giannakos, & Cukurova, 2023). For instance, in their work on embodied learning and maths education using insights from computational models and eye-tracking data, Abrahamson *et al*. (2015) revisit, refine, and elaborate further on some of the seminal claims from Piaget's theory of genetic epistemology (Piaget, 1970) (e.g. his insistence on the role of situated motor-action coordination in the process of reflective abstraction).

### 8. Human-AI Hybrid Intelligence Systems

Going back to three conceptualisations of AI, this leaves us with the last corner of high automation and high human agency, the corner of human cognition being extended with AI in tightly coupled human-AI hybrid intelligence systems (see the AI-HCD Framework in Figure 5 below). The top right corner of the AI-HCD Framework is not covered as much in this paper, mainly because we are yet to see substantial work on this front in AIED. At best, the current complementarity paradigm is to make a better match what humans can do and what AI can do with the problems to be tackled to achieve productivity gains at tasks rather than making humans more intelligent. More commonly, for any given job, we tend to give up our agency to AI to complete a task for us, which in turn is expected to improve the performance of task completion. This inclination to employ AI for task performance is not only appealing but also reflects a natural human propensity for automating processes, which is a trend evident throughout human history (Lubars, & Tan, 2019). However, we must be judicious in selecting the tasks we delegate to AI as the over-reliance on AI could lead to the atrophy of critical competencies in the long term. For instance, there are ongoing attempts to automate qualitative coding processes with the use of large language models, as this can increase the productivity gains in generating labelled qualitative data (Barany *et al.,* 2024). However, on many occasions, qualitative coding is not only done to generate final labelled datasets but also to improve researchers' and practitioners' meaning-making competence through their engagement with the reflective coding process. Similarly for the literature reviews with LLMs, the goal is not only to identify gaps in the literature but to improve one's understanding of the research undertaken on a particular topic. Full automation is unlikely to be valuable with these tasks in the long-term for human competence development unless their use cases are well structured with learning design principles that prioritise human agency.

There is also a significant concern regarding the human-in-the-loop correction of AI-generated content, as this process might lead to convergence towards the AI-generated content and labels rather than critically evaluating them to ground them in our own understanding. Similar to today's society, in which most people converge towards the first suggestion of a Google search engine rather than critically engaging with suggestions to choose the most appropriate one for our goals (Burguet, Caminal, & Ellman, 2015). We might end up in a future where the first generated automated content is taken as "the truth". This is a significant concern for the future of education, therefore, just because we can technically automate a task in education does not always mean that we should. Such significant decisions about automation should be evidence-informed by adopting an approach that involves testing any automation ideas and observing the outcomes in the long term. Otherwise, the rush towards automation may result in humans forfeiting their fundamental cognitive competencies and evolutionary superiority as a species in the future. A potentially wiser perspective would be to highlight the critical importance of maintaining and enhancing our intrinsic intellectual abilities, which have historically conferred upon us distinct advantages for survival and adaptation. This requires alternative conceptualisations of AI.

On the other hand, for AI models that are internalised by humans, the goal of the model would be to "fade away" as humans' competence at the task that is modelled develops through their interaction with the model. At last, extending human cognition for intelligence augmentation in tightly coupled-human AI hybrid intelligence systems would require AI to be a synergistic superstructure built on top of the human intelligence structure, in a way that as the interaction with the tool increases, our competence at the task being modelled would also increase.

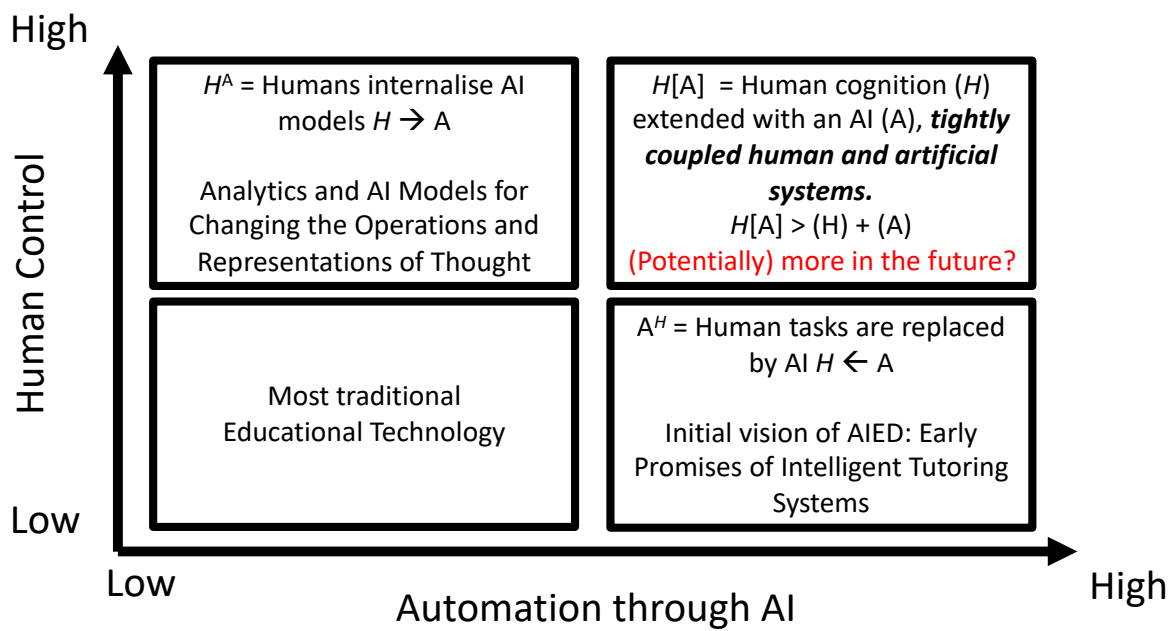

**Figure 5.** The AIED-HCD conceptual framework for human-AI interaction in education for human competence development – Human cognition extended with AI in tightly coupled hybrid intelligence systems.

Our interactions with AI systems are influencing us and we are currently lacking long-term impact studies of these interactions as a research community. In order to be able to achieve human-AI hybrid intelligence systems, we need AI models that are able to interact fluidly with us to shape intents and meanings dynamically. The current AI systems lack the ability to update their models based on interactions with users real-time. Human-AI hybrid intelligence systems would require interactions with AI models in which AI encourages people to shape their own meanings by leveraging the strength of language as a tool for mutual understanding. People very rarely come to an interaction situation with a specific, or fully formed, meaning in mind. Rather such meaning is shaped dynamically through the interaction itself. This requires AI to help humans shape their intents and knowledge in interactions rather than pushing their predictions using language as a query-response approach. Current AI systems' requirement that the meaning should be clearly transferred to an AI for it to generate a relevant output, is a significant limitation against the creation of human-AI hybrid intelligence systems. In such a paradigm, the responsibility of meaning making would also be shared between the two agents, human and AI, rather than humans assuming the entire responsibility for the success of the interaction. In such human-AI hybrid intelligence systems, as humans' interaction with the system increases their competence at the task at hand would also increase. There is an urgent need of research in human-AI interaction in education and I hope that there will be more examples of developing and implementing human-AI hybrid systems in the future.

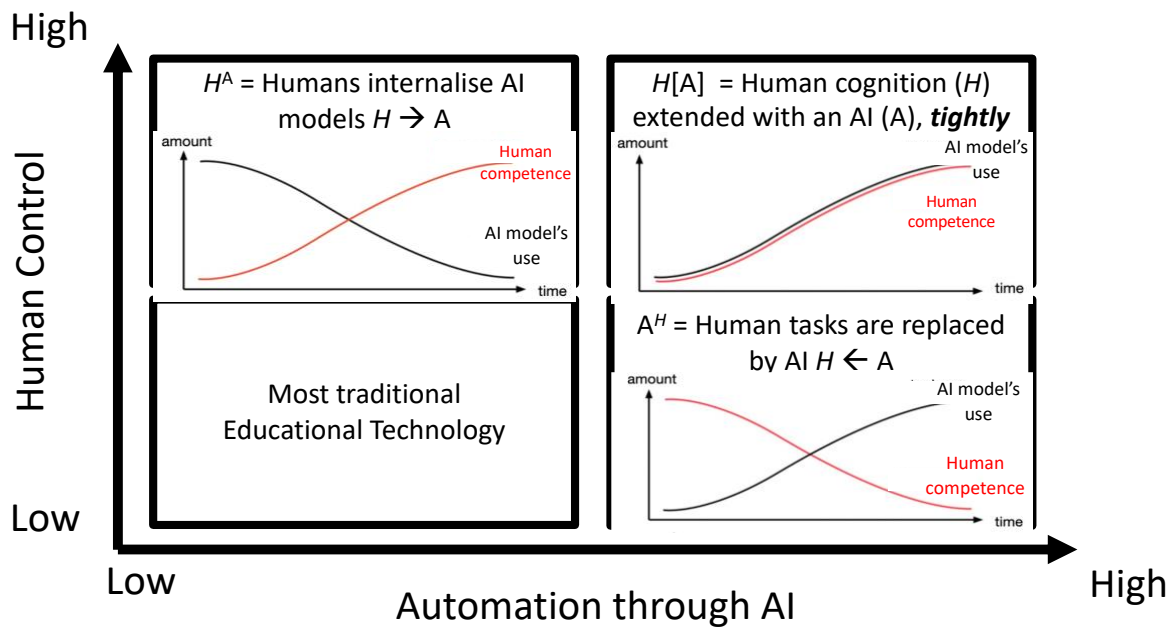

**Figure 6.** The impact of three AI conceptualisations on Human Competence Development in the long term – The AIED-HCD conceptual framework

### 9. AIED is broader than the Design, Development, and Use of AI

Most of the paper has been focusing on the design and development of AI in education and its different conceptualisations including human-AI hybrid intelligence systems. However, the conceptualised human-AI hybrid intelligence system here would also require humans to develop a set of competencies to operationalise such systems and would also require significant innovations in our education systems for them to be integrated. Broadly speaking, in addition to the design, development, and use of AI systems in education, there are two other AI implications for Education that are equally important for such a conceptualisation to make real-world impact in education. One is about educating people about AI and data so that they can learn how to use it safely, effectively, and ethically (see figure 7 below). Another one is work that focuses on how we should think about innovating our education systems so that they are more compatible with, and still relevant to a world heavily influenced by AI (Luckin, & Cukurova, 2019).

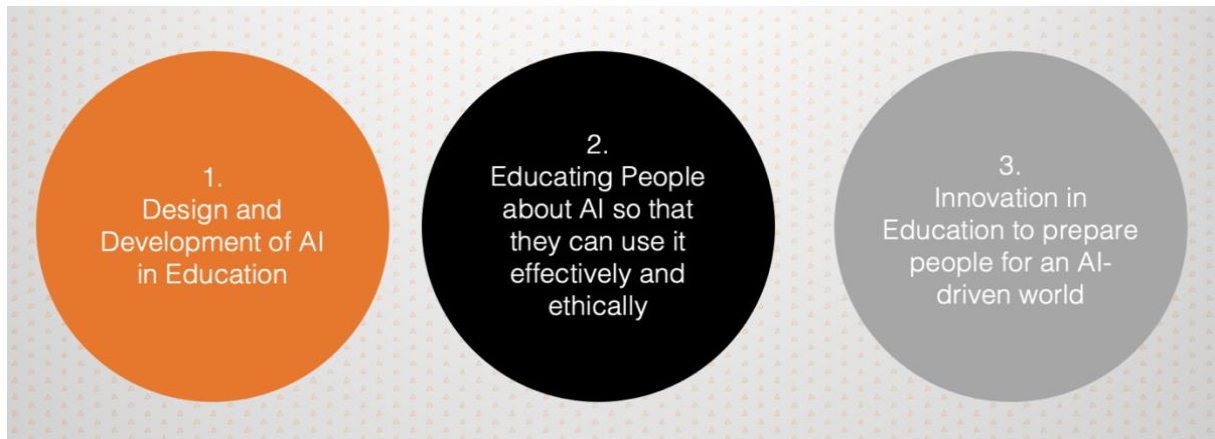

**Figure 7.** Artificial Intelligence's three main implication areas for education.

### 9.1 Educating People about AI

Educating people about AI is indeed about teaching AI to certain groups in population to create AI experts, including AI in our school curricula as well as investing further in our tertiary education institutes to improve capacity to sustain expertise in AI, yet not limited to this. There are many specific research and policymaking initiatives providing guidance in this space (e.g. AI4K12 (Touretzky, Gardner-McCune, Martin, & Seehorn, 2019), UNESCO K-12 AI Curricula).

Perhaps more importantly, it is about helping an overwhelming majority of society, if not everyone, to develop relevant AI competencies so that they can use AI effectively and ethically as part of tightly coupled human-AI hybrid intelligence systems. A recent example in this space is the draft UNESCO AI competency framework[2] for teachers (see Figure 8 below). The key distinctions between AI and previous iterations of ICT tools necessitate the definition of a specific set of competencies for people. For instance, Human-AI hybrid intelligence systems would demand a stronger emphasis on competencies related to human agency, ethical considerations, critical thinking, and human-centred design in human-AI interactions. Training people to develop their competencies on these aspects has utmost importance for the future of human-AI hybrid intelligence systems.

AI competency training is not limited to gaining fundamental AI knowledge, techniques, and skills to apply AI. It is much broader than this. In this version of the UNESCO AI competency framework for teachers, AI techniques and applications is only one of the five main aspects of AI competence. Similar set of competencies would be required from people who are not AI experts but are expected to interact with AI as part of their daily and professional lives. Often time both learning analytics and AIED communities consider the creation of a particular AI tool, model, or dashboard as the end goal of research. However, such an approach is limited for making progress towards a future of human-AI hybrid intelligence. It might be more productive to start considering the tools and models we build as the start of another line of research and practice journey in which end users' competencies are further developed to achieve the expected impact of these solutions. For this, teacher and learner interactions with tools and models should be appropriately scaffolded and supported, rather than assuming that if we left the AI tools in the hands of end users, hybrid intelligence would emerge automatically.

---
[2] https://www.unesco.org/en/digital-education/ai-future-learning/competency-frameworks

| Aspects | Progression | | |
|---|---|---|---|
| | **Acquire** | **Deepen** | **Create** |
| 1. **Human-centred Mindset** | Human agency | Human accountability | AI social responsibility |
| 2. **Ethics of AI** | Ethical principles | Safe and responsible use | Co-creating AI ethical rules |
| 3. **AI Foundations and Applications** | Basic AI techniques and applications | Application skills | Creating with AI |
| 4. **AI Pedagogy** | AI-assisted teaching | AI-pedagogy integration | AI-enhanced pedagogical transformation |
| 5. **AI for Professional Development** | AI enabling lifelong professional learning | AI to enhance organizational learning | AI to support professional transformation |

**Figure 8.** The draft UNESCO AI competency framework for teachers

### 9.2 Innovating Education Systems for an AI-driven World

At last, AI does not only have implications for our education systems that are direct but there are also multiple indirect implications which necessitate innovations in our existing education systems. The future of tightly coupled human-AI hybrid intelligence systems would require significant advancement in our traditional educational systems, particularly with regards to assessment structures.

For instance, although most skills and competencies we are interested in and expect people to develop through education in modern societies are process-driven; often times in traditional education systems these are only evaluated through the outcomes of this competence rather than the process that leads to it. We need to move towards innovations in our assessment structures that encourage process evaluations rather than outcome evaluations only. Recent advancements in LLMs created significant concerns that most student will use them to submit their assignments rather than write essays themselves. This is a valid concern, but it misses the point in the big picture. Essay writing is most often assigned not because the resulting essay has much value to us, but because the process of writing an essay teaches crucial skills to people: regulation of one's own behaviours to engage in a topic, researching a topic, judging claims' accuracy, synthesizing knowledge and expressing it in a clear, coherent and persuasive manner. These skills should be the focus of assessment not *only the final product of what is produced*. This is an example of the kind of innovation we need in our education systems now that AI is ubiquitous.

For example, in addition to traditional content feedback on students' writings, Suraworachet *et al.* (2023) have been providing students with personalised behavioural feedback based on their writing engagement analytics using their data from online word-processing platforms like Word or Google Docs as part of their assessment. When students are engaged with the platform data on time, how much the content is edited, and what content exactly added, are all logged to be able to model their essay writing engagement trends. Based on the analytics of their engagement students are sent feedback with key suggestions on how they can improve their engagement (Suraworachet *et al*., 2022). The feedback has formative suggestions on how they can space their writing practice rather than cramming to complete it all the night before, as well as how previous years' high performing students engaged with their writing assignments.

When the impact of such analytics feedback interventions is evaluated, it was observed that the intervention had limited impact on students who were already doing well in the course, but significantly boosted the engagement and performance of those students who were struggling and were initially predicted to fail (Suraworachet *et al*., 2023). This kind of innovation prioritising process evaluations with analytics is an example of the kind of innovation needed in education systems for assessments to be meaningful if we are to move towards the use of human-AI hybrid intelligent systems in the future.

## 10. Concluding Remarks

Although, most of the current rhetoric is limited to this, AI is more than a set of applied tools we use in education. As presented in the proposed AIED-HCD conceptual framework here (see figures 5 and 6), AI can be conceptualised to externalise, be internalised, or extend human cognition as part of tightly coupled human-AI hybrid intelligence systems. AI can also be a methodology to create computational models of learning as objects to think about learning. While we are doing so, we might notice that some aspects of learning just come through the slow experience of living those learning moments and cannot be fully explained with AI models to be hacked with predictions. Still though, if we take the more modest goal of using AI models for describing these learning processes in more detail and precision, they can provide valuable opportunities for feedback, motivation, awareness, and contributions to theoretical understanding.

Each conceptualisation of AI discussed here might potentially bring some advancements to education, yet they also have significant unintended consequences that should be carefully considered. Our interactions with AI systems are influencing us, and we are currently lacking long-term impact studies of these interactions as a research community. This is a pivotal moment for all of us as scientists and practitioners to envision a future of education that is aligned with our societal values and ensure that AI is used responsibly to achieve this vision. Therefore, this special section on hybrid intelligence has utmost importance to stimulate the line of research on empirical investigations of human-AI interaction in education as well as relevant studies on the implications for improving human competencies in their interactions with AI, and innovating education systems for the future of hybrid intelligence. Based on the emerging evidence in these studies, the future of education with hybrid intelligence systems should be intentional, evidence-informed, and human-centred.

As AIED researchers we finally get the expected attention from society due to recent developments in generative AI, let us not stop questioning who we are as a community, and what we are doing as researchers and practitioners of learning, analytics, and AI. Let us not forget that the purpose of scientific research is only realized to the extent that it helps us understand who we are. Research in AI began as an attempt to understand our own intelligence, its atrophy, its augmentation, or amplification. Which goal are we striving for? When is it acceptable to entrust core cognitive competencies to an AI tool, and when might this pose too great a risk? What are the long-term implications of the tools we are developing on our own competence? We must exercise wisdom in making such decisions and consider alternative conceptualisations of AI in our research and practice. Everything we do as a research community is about humans; it has always been so, and it always should be.

# Acknowledgements

This paper is based on the keynote talk given by the author at the ACM International Conference on Learning Analytics & Knowledge (LAK) 2024 in Kyoto, Japan. The ideas presented here are shaped during the years of discussions and debates with many colleagues to whom I am grateful. I would like to thank Prof. Sanna Järvelä, Prof. Guoying Zhao, Dr. Andy Nguyen, and Dr. Haoyu Chen for their valuable comments during the review process of this manuscript.